\documentclass[twocolumn,prl,aps,superscriptaddress,nobalancelastpage]{revtex4-1}
\usepackage{graphicx,bm}
\begin{document}

\title{Nesting of electron and hole Fermi surfaces in non-superconducting BaFe$_2$P$_2$}

\author{B.J. Arnold}
\affiliation{H. H. Wills Physics Laboratory, University of Bristol, Tyndall Avenue, Bristol, BS8 1TL, United Kingdom.}
\author{S. Kasahara}
\affiliation{Research Center for Low Temperature and Materials Sciences, Kyoto University, Sakyo-ku, Kyoto 606-8501,
Japan.}
\author{A.I. Coldea}
\affiliation{Clarendon Laboratory, Department of Physics, University of Oxford, Oxford OX1 3PU, United Kingdom.}
\affiliation{H. H. Wills Physics Laboratory, University of Bristol, Tyndall Avenue, Bristol, BS8 1TL, United Kingdom.}
\author{T. Terashima}
\affiliation{Research Center for Low Temperature and Materials Sciences, Kyoto University, Sakyo-ku, Kyoto 606-8501,
Japan.}
\author{Y.~Matsuda}
\affiliation{Department of Physics, Kyoto University, Sakyo-ku, Kyoto 606-8502, Japan.}
\author{T.~Shibauchi}
\affiliation{Department of Physics, Kyoto University, Sakyo-ku, Kyoto 606-8502, Japan.}
\author{A. Carrington}
\affiliation{H. H. Wills Physics Laboratory, University of Bristol, Tyndall Avenue, Bristol, BS8 1TL, United Kingdom.}

\begin{abstract}
We report detailed measurements of the de Haas-van Alphen (dHvA) effect in BaFe$_2$P$_2$ which is the end member of the
superconducting series, BaFe$_2$(As$_{1-x}$P$_x$)$_2$. Using high purity samples we are able to observe dHvA
oscillations from all the sheets of Fermi surface and hence build up a complete detailed picture of its structure.  The
results show the existence of a highly warped section of hole surface which may be the origin of the nodal gap
structure in this series.  Importantly, we find that, even for this non-superconducting end member, one of the hole
surfaces almost exactly nests with the inner electron surface.  This suggests that improved geometric nesting does not
drive the increase in pairing strength with decreasing $x$ in this system.
\end{abstract}

\maketitle

A common feature that links together the various families of iron-based superconductors is their unusual electronic
structure, which consists of almost nested, quasi two dimensional electron and hole Fermi surfaces (FS). This quasi
nesting produces strong peaks in the real and imaginary parts of the susceptibility (both bare and interacting), which
in many theories \cite{MazinSJD08,KurokiOAUTKA08} drives the superconductivity via a spin-fluctuation mechanism.
However, the relative importance of geometric nesting and the many-body interactions ($U$ and $J$ in Hubbard type
models) remains under debate.  Studies of how the FS evolves and correlates with the changing material properties over
the phase diagram are therefore of great importance.

 Although, quasi nested electron and hole bands are a common
feature of calculated band-structures of these compounds, the exact topology and size of the FS is very difficult to
compute from first principles. There are usually four or five bands which cross the Fermi energy, giving rise to small
Fermi pockets each of which fills $\sim 10-20$\% of the Brillouin zone volume. The small size of these pockets means
their size and topology are very sensitive to small shifts in the band energies, which can result from small structural
changes or from interactions with spin-fluctuations which scatter electrons between the electron and hole sheets
\cite{Ortenzi09}. Hence, it is important to experimentally determine the full three dimensional FS so that
 quantitative tests of the theory can be made. Measurements of the de Haas-van Alphen (dHvA) effect provide a
very accurate way of measuring the bulk FS.

\begin{figure}
\center
\includegraphics*[width=8cm]{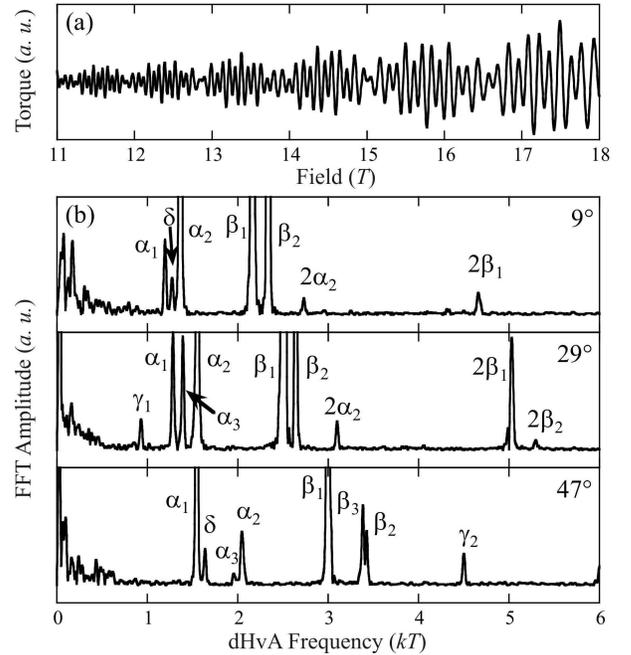}
\caption{(a) An example torque versus field trace with background subtracted (b) Fast Fourier transforms of the $\tau(H)$ data (from 6-18\,T)at
various angles showing the distinct dHvA peaks. Harmonics are identified with the prefix 2. All data was taken at base temperature $T\simeq 0.35$\,K. Note the peaks
below 0.5\,kT are due to noise and features in the background cantilever magnetoresistance.} \label{Fig:Rawdata}
\end{figure}

A particularly useful system for studying the influence of FS topology on the ground state of the iron-based
superconductors is BaFe$_2$(As$_{1-x}$P$_x$)$_2$.  Here, the N\'eel temperature of spin density wave ground state of
the end member BaFe$_2$As$_2$ is driven to zero and superconductivity emerges with a maximum $T_c=30$\,K, as As is
substituted with the isovalent element P. So unlike charge doped iron-pnictide systems, (e.g.,
Ba$_{1-x}$K$_x$Fe$_2$As$_2$), BaFe$_2$(As$_{1-x}$P$_x$)$_2$ is expected to remain perfectly compensated (equal volumes
of electron and hole pockets) for all $x$.  Minimal disorder is introduced in the Fe plane by these substitutions so
that dHvA oscillations can be observed over a wide range of $x$ ($0.42<x\leq 1$) \cite{Shishido10}.

\begin{figure*}
\center
\includegraphics*[width=16cm]{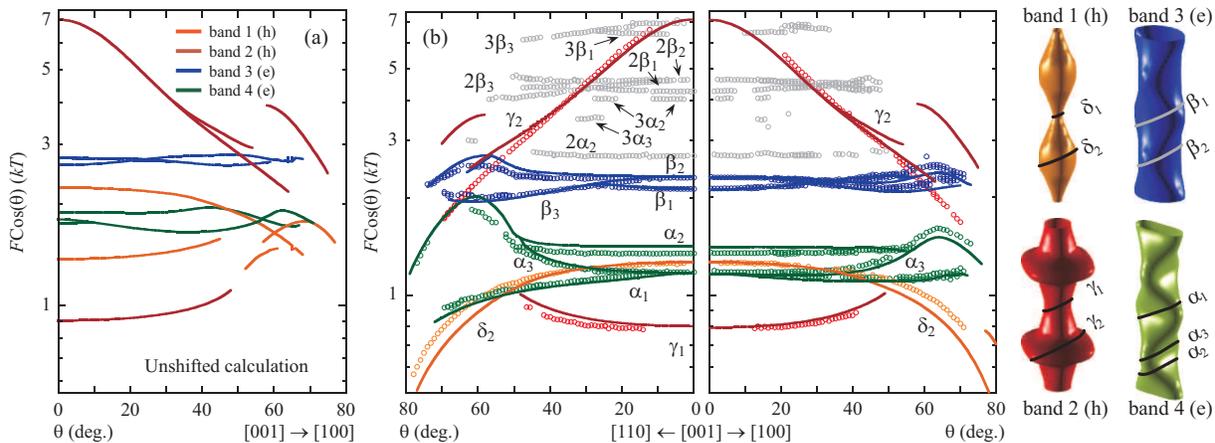}
\caption{(color online) dHvA frequencies multiplied by cosine of the rotation angle versus angle. (a) Shows the
calculations for the unshifted bands rotating from [001] towards [100].  In (b) the symbols show experimental data for
rotations from [001] to both [100] and [110] whereas the solid lines show the calculations for the energy shifted bands
as described in the text (note that the predicted orbit $\delta_1$ which has $F\simeq 100$\,T at $\theta=0^\circ$ is
not shown) \cite{rotnote}. (c) Shows the origin of the orbits on the calculated electron (e) and hole (h) FS sheets at
an angle of $\theta=20^\circ$ (towards [110]).} \label{Fig:Rot}
\end{figure*}

Structurally, increasing $x$ results in a decrease in the $c/a$ ratio and a movement of the pnictogen atom towards the
Fe plane.  Density functional theory (DFT) calculations based on the local density approximation (and its variants)
predict that throughout the series the electron sheets change very little, but for the hole sheets there are much
greater changes \cite{Shishido10}.   BaFe$_2$As$_2$ has three quasi 2D hole sheets but for BaFe$_2$P$_2$ one of these
disappears and one of the hole sheets develops pronounced warping close to the top of the zone (dumbbell, band 2 in
Fig.\ \ref{Fig:Rot}). A notable property of the BaFe$_2$(As$_{1-x}$P$_x$)$_2$ series is that there is clear evidence
from multiple probes \cite{Hashimoto0907.4399,Nakai10a} that there are line nodes in its superconducting energy gap. It
has been proposed theoretically that these nodes are located on the dumbbell hole surface \cite{Suzuki1010.3542}
although there is experimental evidence that rather points to the nodes being located on the electron sheets
\cite{Hashimoto10,Yamashita1103.0885}. Establishing a clear connection between the unusual gap structure of
BaFe$_2$(As$_{1-x}$P$_x$)$_2$ and its FS topology would be a strong confirmation that the pairing mechanism is indeed
mediated by spin-fluctuations in these materials.

In previous studies, dHvA signals from \textit{all} of the FS sheets were not observed and so the complete
 topology could not be determined. In particular, Shishido \textit{et al.} \cite{Shishido10} only observed
signals from the electron sheets.  Following this, Analytis \textit{et al.} \cite{Analytis10} reported measurements for
$x=0.63$ where orbits on the smaller hole sheet were observed and were found to correspond to almost the same
crosssectional area as the smaller electron pocket.

Here we report new measurements of the dHvA effect for the end member BaFe$_2$P$_2$.   The much higher purity level of
our new samples allows us to observe dHvA oscillations originating from all the FS sheets and hence we are able to make
a precise determination of the complete FS.  A surprising aspect of the results is that they show that even for this
non-superconducting (and non-magnetically ordered) composition, one of the hole sheets is almost perfectly nested with
the (smaller) electron FS, similar to that found for the superconducting composition $x=0.63$ \cite{Analytis10}. This
suggests that nesting is probably a necessary but not sufficient ingredient for superconductivity.

Samples of BaFe$_2$P$_2$ were grown as described in Ref.\ \cite{Kasahara10}.   The lattice parameters and the crystal
orientation was determined by x-ray diffraction.  The best results were obtained on a sample with dimensions $50\times
75 \times 20 \mu$m, which was mounted on micro piezo resistive cantilever in order to measure the magnetic torque. The
torque sensor was fixed to a rotating probe, inside a $^3$He cryostat equipped with a superconducting magnet.  The
orientation of the sample on the cantilever was determined by comparing optical images of the mounted sample with
images taken on the x-ray mount. We estimate the accuracy of the in-plane alignment $\phi$ is around $\pm 5^\circ$.

\begin{figure*}
\center
\includegraphics*[width=16cm]{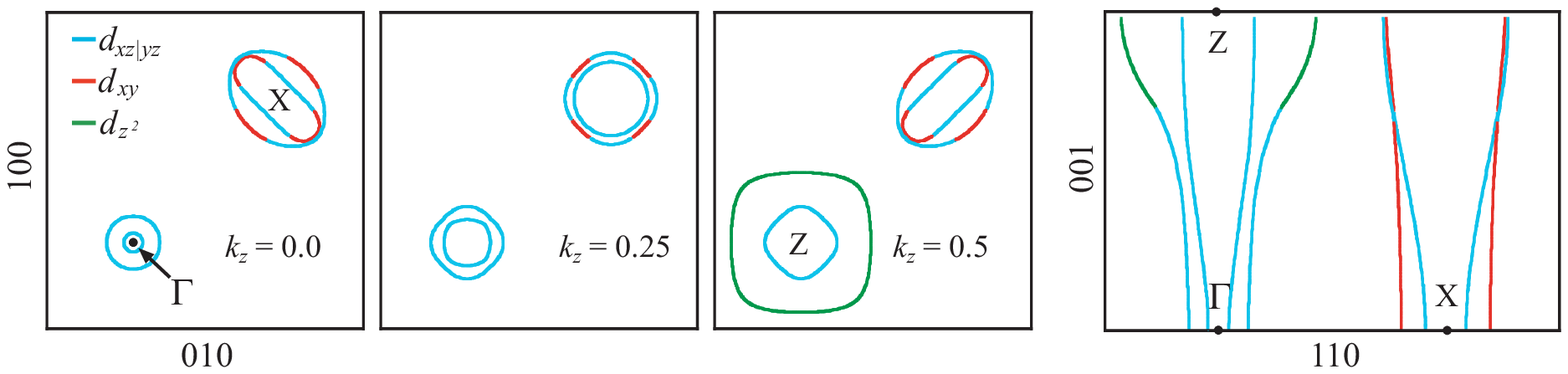}
\caption{(color online) Crosssections of the determined Fermi surface of BaFe$_2$P$_2$ in: (a) the ab plane at three
different $k_z$ values (quoted in units of $c^*$), (b) a $c$-axis cut along the [110] direction and through the
$\Gamma$ point.  The largest band character element at each $\bm{k}$-point is indicated \cite{shiftnote}.}
\label{Fig:FSplots}
\end{figure*}

Example torque $\tau$ versus field $H$ data is shown in Fig.\ \ref{Fig:Rawdata}.  In this sample, the oscillations are
visible down to fields as low as 5\,T at some angles.  By taking fast Fourier transforms of the $\tau(B^{-1})$ data we
identified 8 different dHvA orbits.  As many of these peaks are very close in frequency they can only be clearly
identified by performing long field sweeps (typically between 6 and 18\,T) with small (1 degree) angular increments. In
this way the various crossing branches can be easily identified.  Complete rotations were performed in two principle
directions, namely from $H$ perpendicular to the Fe plane [001] towards either the [100] or [110] directions.  In both
cases  we define the rotation angle $\theta=0$ for $H\|$[001].

For a perfectly two dimensional FS sheet, the dHvA frequency $F\propto 1/\cos(\theta)$, hence for quasi-two dimensional
systems the dHvA data are best appreciated by plotting $F\cos \theta$ versus $\theta$ as in Fig.\ \ref{Fig:Rot}.  In
such a plot, a local minimum in the FS crosssection (neck) gives rise to a branch where $F\cos\theta$ increases with
$\theta$, whereas for a local maximum (belly)  $F\cos\theta$ decreases with $\theta$.

In order to make a more precise analysis of the results we compare the data with predicted extremal orbits derived from
bandstructure calculations.  These density functional calculations were done using an augmented plane wave plus local
orbits scheme as implemented in the Wien2K package \cite{wien2k,calcnote}.  The calculated extremal orbits as function
of the field angle for a [001] to [100] rotation are shown in Fig.\ \ref{Fig:Rot}(a).  Although this calculation
reproduces the general features of the experiment, the exact topology and FS size is not well reproduced.  As we have
found in other pnictide systems \cite{ColdeaFCABCEFHM08,Analytis09} the calculation can be brought into much better
agreement with experiment by performing rigid ($\bm{k}$-independent) shifts of the band energies.  The two electron
bands (4 and 3) are easily identified as the $\alpha$ and $\beta$ branches respectively and these are brought into
almost exact agreement by shifting the energies of band 4 up by 68\, meV and band 3 up by 58\, meV.  These surfaces
have quite a complicated topology which is well reproduced in the calculations [see Fig.\ \ref{Fig:Rot}(b)]. Note the
large differences between the rotation toward [100] and [110]. The large (highest fundamental frequency) orbit
$\gamma_2$ originates from the highly warped part of the band 2 hole `dumbbell' surface centered around the top of the
zone. The calculations reproduce this branch without \emph{any} applied energy shift. The curvature of $F\cos(\theta)$
for the $\delta_2$ branch clearly identifies it as a maximum (belly) crosssection and so it must originate from band 1
as this is the only remain unidentified \textit{maximal} orbit. Shifting the energy of this band down by 113\,meV gives
an almost exact match. With this shift, the minimal crosssection $\delta_1$ of this band would give a dHvA frequency
around 100\,T.  We do not see conclusive evidence for this orbit probably because its expected weak signal is likely
buried in the low frequency cantilever noise (see Fig.\ \ref{Fig:Rawdata}). Finally, we can identify the last branch
$\gamma_1$, as the minimum of band 2. Although no energy shift was required to explain the maximum of this band, we
require a shift down of 52\,meV to fit the minimum. With these energy shifts, the FS is accurately determined. As a
check we calculate the number of electron and holes in each shifted FS sheet and find that, as expected, they are equal
to within $0.017$ electrons per unit cell.

It has been suggested that physically these shifts result from a spin fluctuation interaction \cite{Ortenzi09}.
Scattering of electrons by the spin-fluctuations between the electron and hole surfaces causes both surfaces to shrink
in volume and hence the energy shifts are in opposite directions.  This is supported by results for CaFe$_2$P$_2$
\cite{Coldea09}, which has a non-nested FS similar to the collapsed tetragonal, high pressure phase of CaFe$_2$As$_2$,
and where the DFT calculations are in excellent agreement with experiment without any shifting. Further, for
BaFe$_2$(As$_{1-x}$P$_x$)$_2$ it was found that the size of energy shifts varied approximately linearly with $x$,
becoming larger as the SDW phase transition is approached \cite{Shishido10} and the spin-fluctuations grow in
magnitude. Our results here also support this interpretation, because we find that most sections of the FS of
BaFe$_2$P$_2$ are close to being geometrically nested and are smaller compared to the DFT calculation whereas the
highly warped part of band 2  dumbbell ($\gamma_2$ orbit) is far from nesting and remains unchanged. This supports the
suggestion of Suzuki \textit{et al.} \cite{Suzuki1010.3542} that this part of the Fermi surface does not contribute
strongly to the pairing and is therefore may be susceptible to gap node formation.

\begin{table}
    \caption{Measured effective masses $m^*$ (uncertainty $\sim$2\%) along with mass enhancements $m^*/m_b$ calculated using the bandstructure values
    ($m_b$) from the energy shifted bands. The mean free paths $\ell$ are also reported \cite{supp}.}
    \begin{tabular}{llllllll}
    \hline \hline
        & $\gamma_1$ & $\gamma_2$& $\delta$ & $\alpha_1$ & $\alpha_2$ &  $\beta_1$ & $\beta_2$  \\
    \hline
    $\theta$ & 28$^\circ$ & 46$^\circ$&12$^\circ$&12$^\circ$&12$^\circ$&12$^\circ$&12$^\circ$\\
    $m^*/m_e$ & 2.30 & 3.32& 1.64 & 1.75 & 1.58 & 1.65 & 1.54  \\
    $m^*/m_b$ & 1.59& 1.59 & 1.80 & 1.82 & 1.88 & 1.68 & 1.81  \\
    $\ell$(nm)& 40  &150 & 40& 190& 200& 170&160\\
\hline
\end{tabular}
\label{Table:mass}
\end{table}

The masses $m^*$ of the quasiparticles on each of the extremal orbits were determined by measuring the temperature
dependence of the dHvA oscillation amplitude \cite{supp}.  Fits to this data to determine $m^*$  are reported in Table.
\ref{Table:mass}, along with the masses $m_b$ calculated from the shifted band-structure bands for the same orbits and
at the same field angles.  The mass enhancements $m^*/m_b$ are all relatively similar and small ranging from 1.6 to
1.9.  Note the values for the $\beta$ orbits are smaller than the value give in Ref.\ \cite{Shishido10} but consistent
within the error.

An interesting feature of the results is that they show that for this non-superconducting end member of the
BaFe$_2$(As$_{1-x}$P$_x$)$_2$ series, one of the electron and hole sheets are very well nested.  The closeness of the
crosssectional areas of the inner electron and hole FS is evident from the rotation plots shown in Fig.\ \ref{Fig:Rot}.
The fact that these areas are so close explains why they were not observed in previous studies, where the dHvA
oscillations were not observable over such a wide range of inverse field as that were here.  In Fig.\ \ref{Fig:FSplots}
we show sections through the determined FS.  Although the in-plane area of the electron sheet FS does not change
strongly with $k_z$, its shape does.  Because of the symmetry of the body centered tetragonal unit cell the
crosssection at the zone center is exactly the same as at the zone top but rotated by $\pi/2$. The best match, where
the inner electron sheet is almost perfectly circular is half way between these points at $k_z=0.25$. Furthermore the
band character of these two sheets is also very similar; they both have predominately Fe $d_{xz/yz}$ character.

Comparing our data with that of Analytis \emph{et al.}\cite{Analytis10} for the superconducting ($T_c=7$\,K)
composition $x=0.63$ suggests that besides the above mentioned shrinking of the overall size of the Fermi surface there
is remarkably little change in the overall topology with $x$ (for $0.63\leq x\leq 1$).  A close comparison between the
two data sets \cite{supp} shows that the $F(\theta)$ data may be almost perfectly superimposed by simply scaling the
mean frequencies of the electron bands. This suggests that the $c$-axis warping of the electron and hole sheets remain
remarkably constant as a function of P content $x$.  The anomalous downturn in the frequency of the electron $\alpha$
branch between $\theta=40^\circ$ and $50^\circ$ identified in Ref.\ \cite{Analytis10} was correctly assigned to a
crossing of the hole and electron orbit frequencies.  However, the higher scattering in these doped samples probably
prevented the observation of the dumbbell ($\gamma_2$) orbit. In Ref.\ \cite{Analytis10} the $\gamma$ branch was
assigned to the minimum of band 1.  However, as the size and dispersion of this branch is almost identical to the
$\gamma_1$ branch in the present work we suggest that this assignment was incorrect, and it is instead the minimum of
band 2 as in the present work. This effects somewhat the fitted band structure of Ref.\ \cite{Analytis10} but does not
change the conclusion that the electron and hole bands are well nested for $x=0.63$.

Combined with the present results it appears then that rather than the bands become much better nested as $x$
approaches that optimal for superconductivity, the \emph{relative} size of the electron and hole sheets remain
remarkably constant.   As the electron sheets shrink with decreasing $x$ the hole bands shrink by a similar amount
without any major change in topology. This is not expected from conventional bandstructure calculations, where the hole
band topology changes very dramatically as $x$ is decreased.  It is therefore likely that the dumbbell shape of the
larger hole band remains for the superconducting compositions. Recent photoemission results for the $x=0.38$
composition \cite{Yoshida1008.2080} indicate the FS topology of this close to optimal $T_c$ composition is indeed close
to that observed here for the end member.

In summary, our results determine, in very high detail, the structure of the FS of the end member of the
BaFe$_2$(As$_{1-x}$P$_x$)$_2$ series.  Together with the known shrinking of the average size of the electron pockets
and invoking the expected compensated nature of the FS, it is now possible to estimate the topology of both the
electron and hole surface across the phase diagram.  An important aspect of the results is that they show that one of
the hole and electron FS are well nested even for this non-superconducting (and non magnetic) composition. This
suggests that the increase in pairing susceptibility as $x$ is decreased is caused by increases in the many body
interaction parameters $U$ and $J$ rather than any effect of improved geometric nesting.

We thank Mairi Haddow for assistance with the x-ray diffraction measurements and Ed Yelland for the use of computer
code. This work was supported by the UK EPSRC  and by KAKENHI from JSPS .

\bibliographystyle{aps5etal}
\bibliography{BaFe2P2dHvA}

\begin{thebibliography}{20}
\bibitem[{rot(????)}]{rotnote}
\bibinfo{note}{For the calculations of the [001]-[100] rotation we used an
  in-plane angle of $\phi=10^\circ$ as this gave the best fit whereas for the
  [001]-[110] we used exactly $\phi=45^\circ$.}
\bibitem[{shi(????)}]{shiftnote}
\bibinfo{note}{For the outer hole band we smoothly interpolated between the
  shifted and unshifted Fermi surfaces to give the correct crosssections for
  the neck and belly orbits of this surface for this figure.}
\end{thebibliography}

\end{document}